# Spatial-Spectral Complexity in Kerr Beam Self-Cleaning

M. Labaz[1, *] and P. Sidorenko[2].

[1] Department of Physics and Solid-State Institute, Technion – Israel Institute of Technology, 32000 Haifa, Israel
[2] Department of Electrical and Computer Engineering and Solid-State Institute, Technion – Israel Institute of Technology, 32000 Haifa, Israel
*Corresponding author: l.moshe@campus.technion.ac.il

**We report on a comprehensive experimental investigation into the spatial-spectral complexity of the laser beam during Kerr-induced beam self-cleaning in graded-index multimode fibers. We demonstrate the self-cleaning of beams using both transform-limited and chirped femtosecond pulses. By utilizing the spectrally resolved imaging technique, we examine variations in beam homogeneity during the beam cleanup process and reveal correlations observed among spatial beam profiles at different wavelengths for the various cleaned pulses. Our results significantly advance our understanding of Kerr-induced self-cleaning with chirped ultrafast pulses and offer new possibilities for diverse applications.** © 2024 Optica Publishing Group

http://dx.doi.org/xxxxxx/xxxxxxx

In recent years, a significant and renewed focus has been on researching nonlinear waves in multimode optical fibers (MMFs). The inclusion of spatial degrees of freedom in MMFs allows for new possibilities in the exploration of spatiotemporal nonlinear complexities [1–3]. The spatiotemporal pulse propagation in MMF results in various fascinating phenomena, including multimode soliton [4], geometric parametric instability [5], spatiotemporal mode-locking [6], multimode supercontinuum generation [7], and more.

Kerr Beam Self-Cleaning (KBSC) is a fascinating phenomenon observed in multimode fibers. In this phenomenon, the Kerr-induced nonlinear mode coupling causes a speckle pattern to transform into a robust bell-shaped profile as the power levels rise. Initially, KBSC was observed in graded-index (GRIN) MMF under normal dispersion using both nanosecond [8] and femtosecond [9] pulses. Subsequent studies have shown the occurrence of KBSC in different types of active and passive fibers, encompassing both normal and anomalous dispersion regimes [1]. A thermodynamic theory has recently been proposed to elucidate the underlying mechanisms behind the KBSC phenomenon [10,11]. This theory describes the multimode wave phenomenon of KBSC as a thermalization of a dilute photon gas governed by the principles of statistical mechanics. A notable finding of the thermodynamic theory is that the modal distribution in the thermalized state aligns with the Rayleigh-Jeans distribution, a validation confirmed through experimental studies [12]. Although the thermodynamic theory assumes continuous waves, all conducted experiments involve pulses. Nonetheless, there is a very good agreement between the theory and experimental results, even when using a broadband femtosecond pulses [12].

Kerr beam cleanup holds significant potential for applications involving high-power lasers and pulse-delivery systems, but there is limited understanding of its spatiotemporal behavior. Recent work revealed nontrivial spatiotemporal dynamics of ns pulses in the process of KBSC [13]. Leventoux et al. investigated KBSC using ns pulses and demonstrated non-uniformity of the spatial beam profile over the temporal shape of the pulse. Additionally, recent studies have shown that beam cleaning with sub ns pulses involves nonlinear polarization rotation and a notable enhancement in the degree of linear polarization [14]. These studies unveiled a complex behavior of KBSC with ns and sub ns pulses in both temporal and polarization degrees of freedom. Hence, it becomes crucial to thoroughly investigate and comprehend the spatiotemporal dynamics of broadband and femtosecond pulses during the process of KBSC.

In this work, we present the experimental investigation of spectrally resolved KBSC with femtosecond pulses. In addition, we demonstrate self-cleaning using chirped femtosecond pulses in GRIN multimode fiber. We employ a spectrally resolved imaging technique [15] to investigate the intricate nonlinear dynamics involved in KBSC. Utilizing this imaging technique, we can uncover the spatial-spectral complexities (SSC) of the cleaned beam, which would otherwise remain undetectable. Unveiling the SSC allows us to get a deeper look at the phenomena that occur during the KBSC process and comprehend its impact on the beam at different wavelengths. These results significantly advance our understanding of Kerr-induced self-cleaning with broadband ultrafast pulses and offer new possibilities for diverse applications.

In experiments, we launch transform-limited or chirped ultrafast pulses into 30 cm of multimode GRIN fiber (GIF625, 62.5 μm core diameter) and measure the spectrally resolved output beam profile at different power levels. Our laser (Pharos, Light Conversion) delivers 175 fs transform-limited pulses at 1030 nm central wavelength. Utilizing an internal grating compressor within the laser system, we adjust the pulses' temporal duration resulting in positive or negative chirping effects. The experimental setup is depicted in Figure 1. The laser beam is coupled into a fiber, and the excitation conditions are controlled by laterally shifting the input fiber face with respect to the laser beam. We imaged the output face of the fiber with a lens and split the image into two planes. One plane holds a CCD camera, solely used for alignment and comparison purposes. The other image plane contains a single-mode fiber (SMF) mounted on a two-dimensional piezo stage. The output of the SMF is connected to a high-resolution spectrometer. The piezo stage is used to scan the imaging plane; at each spatial point, we record the optical power spectrum. The spatial-spectral information is collected, and stored in a 3D array with two spatial and one spectral dimension. An example of such an array is graphically represented in Figure 2(a), which shows the pulse intensity with its integrated projections on different plans. The information is then used to resolve the spatial-spectral dynamics. A calibration measurement illustrated in Fig – 2 (b) shows a comparison between an image taken with the CCD camera (left) and a spectrally integrated 3D array (the calibration procedure is described in Supplement 1 S1). Figure 2(c) illustrates the spatially integrated spectrum and sampled beam profiles at various wavelengths (derived from the same data as in Fig 2(a)).

We investigate and compare the KBSC phenomenon in three experiments: (1) with a transform-limited pulse, (2) with a down-chirped pulse, and (3) with an up-chirped pulse. All pulses have the same spectral Root Mean Square (RMS) bandwidth: ~14 nm, but the chirped pulses are stretched to 5 ps with positive or negative Group Delay Dispersion (GDD) while the transform-limited pulse has 175 fs temporal width. Figure 3 shows the spectrally integrated images of the output beam for increasing power in the experiments. We maintained the same excitation conditions throughout the experiments for all three cases. This is evident from the similarity in the beam profile at low power levels for both chirped and transform-limited pulses (left column in Fig 3). Subsequently, we increased the power in each experiment until observing the energy converging towards the center of the waveguide, resulting in a bell-shaped beam profile. The KBSC can be observed with 63.8 KW for chirped pulses and 47.8 KW for the transform-limited pulses. Importantly, it is difficult to distinguish between the chirped and transform-limited cases when observing spectrally integrated beam profiles at the highest power (Fig. 3 right-most column). However, employing spectrally resolved measurements unveils the intricate nature of the KBSC phenomena.

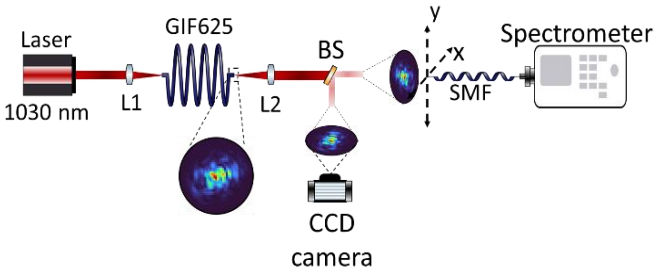

**Fig. 1.** Experimental set-up for Spatial-Spectral (SS) beam mapping.

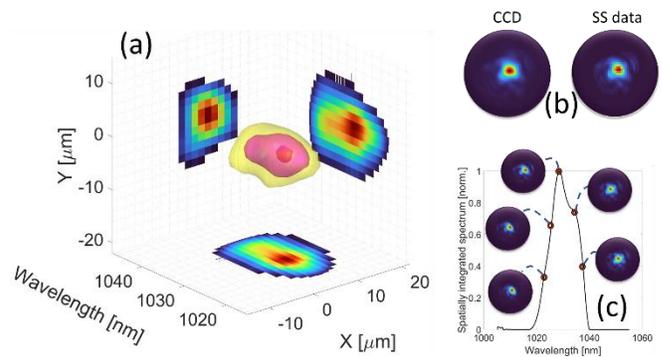

**Fig. 2.** (a) 3D mapping of the SS measurements with integrated projections on the different axes (b) Beam profile obtained by the CCD camera (left) compared with the spectrally integrated data obtained by the SS measurement (right) (c) Spatially integrated spectrum with sampled spatial beam profiles at different wavelengths.

Figure 4 show spatially integrated spectra and beam profiles at the specific wavelengths (indicated by red dots) for the highest power for the down-chirped (DC), up-chirped (UC), and transform-limited (TL) pulses, respectively (the complete data that include all power levels is presented in Supplement 1 S2). The observed distinctions are apparent in the spatially integrated spectral shapes (depicted by the black curves in Fig 4). Notably, the TL pulse exhibits the widest bandwidth due to self-phase modulation induced spectral broadening. The DC pulse displays the narrowest spectrum due to the spectral narrowing. A different behavior of wavelength resolved spatial profiles emerges among the variously chirped pulses. In TL pulses Fig. 4(c), the KBSC primarily manifests at the spectral edges, transitioning to a more speckle pattern towards the central wavelength. Conversely, in DC pulses Fig. 4(a), pronounced KBSC occurs while approaching the spectral center, with speckle patterns emerging at the band edges. Notably, the UC pulses Fig. 4(b) demonstrate the most uniformly distributed patterns across its spectral shape. From the experimental results presented in Fig. 4 it is clear that KBSC is not homogeneous across different wavelengths of the pulse.

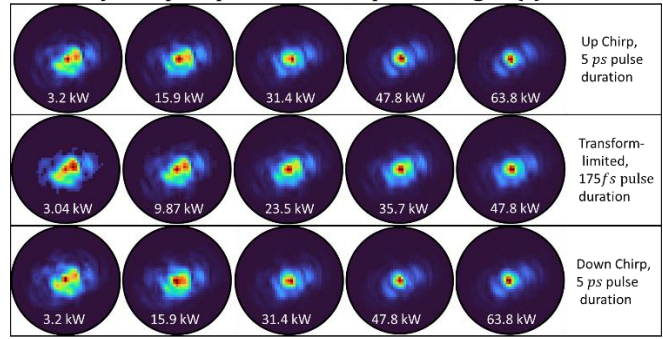

**Fig. 3.** Spectrally integrated data obtained using the SS measurements for three different chirped pulses with increased powers reaching KBSC.

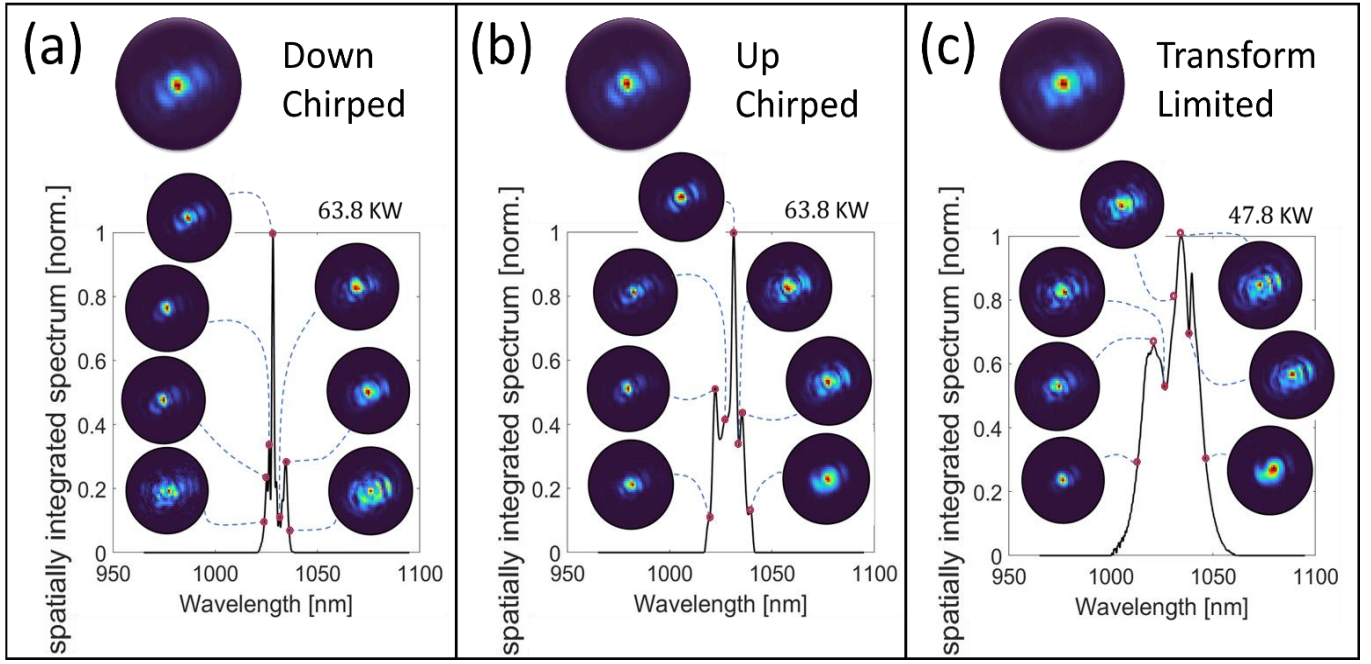

**Fig. 4.** Spatial-spectral results for the three experiments. Down-chirped pulse (2), up-chirped pulse (b) and transform-limited pulse (c). The upper image in each column corresponds to spectrally integrated beam profile.

To quantify the spatial-spectral homogeneity, we employ the commonly used metric [16]

$$V = \frac{1}{N}\sum_{x_i y_j}^{\sqrt{N}} \frac{\left(\int \sqrt{\bar{I} \cdot I_{x_i y_j}(\lambda)}\, d\lambda\right)^2}{\int \bar{I} d\lambda \int I_{x_i y_j}(\lambda)\, d\lambda}. \quad (1)$$

Where – N is the number of measured spatial pixels, $\bar{I}$ is the averaged power spectrum and - $I_{x_i y_j}(\lambda)$ is the power spectrum at the $x_i - y_j$ location. Figure 5 (a) illustrates the homogeneity for chirped and transform-limited pulses. As the power increases, the homogeneity decreases for all cases, but it reaches a saturation point for the down-chirped pulse (yellow curve in Fig. 5(a)) and declines more rapidly for the transform-limited pulse. To some extent, this behavior is inversely correlated with the spatially integrated RMS spectral bandwidth depicted in Fig. 5 (b). The RMS spectral bandwidth increases rapidly for transform-limited pulses, gradually for up-chirped pulses, and decreases for down-chirped pulses. It appears that the beam homogeneity of the up-chirped and transform-limited pulses decreases monotonically. However, the down-chirped pulse behaves differently, which could be due to its decrease in bandwidth compared to the others. Another noteworthy observation is the correlation between the spectral bandwidth changes rate and homogeneity. Specifically, the transform-limited pulse experiences the most rapid changes in bandwidth and the quickest decrease in homogeneity. On the other hand, the up-chirped pulse displays slower changes in both aspects. Regarding the down-chirped pulse, while the bandwidth does decrease, the rate of change falls somewhere in between, leading to a decrease in homogeneity up to a certain level of saturation. This poses the question of how down-chirped pulses behave at higher power levels and with different excitation conditions. Those questions will be investigated in follow up work. As we examine the pulses shown in Figure 4, we can observe that they produce distinct spatial profiles across the spectrum. We employed the Pearson product-moment correlation coefficient [17] to better understand the correlations between these patterns. This allowed us to gain a more insight into the spatial patterns. We use correlations $r_{\lambda_{kl}}$ defined by:

$$r_{\lambda_{kl}} = \frac{\sum_{x_i y_j}\left(I_{x_i y_j \lambda_k} - \bar{I}_{\lambda_k}\right)\left(I_{x_i y_j \lambda_l} - \bar{I}_{\lambda_l}\right)}{\sqrt{\left[\sum_{x_i y_j}\left(I_{x_i y_j \lambda_k} - \bar{I}_{\lambda_k}\right)^2\right]\left[\sum_{x_i y_j}\left(I_{x_i y_j \lambda_l} - \bar{I}_{\lambda_l}\right)^2\right]}}. \quad (2)$$

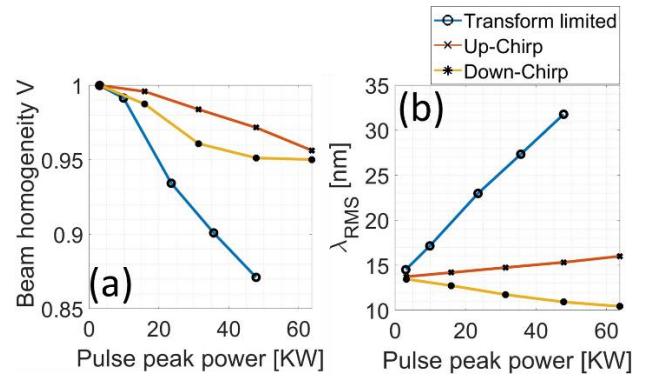

**Fig. 5.** Beam homogeneity alongside with spectral bandwidth for increasing peak powers. (a) Beam homogeneity VS pulse peak power (b) Pulse bandwidth VS pulse peak power.

Where - $I_{x_i y_j \lambda_k}$ is the measured intensity at spatial point $(x_i, y_j)$ and wavelength $\lambda_k$. The $\bar{I}_{\lambda_k}$ represent the spatially averaged intensity at wavelength $\lambda_k$. The correlations of the high-power data (the same data that is presented in Fig. 4) are illustrated in Fig. 6 (b)-(d). For comparison Fig. 6(a) shows the correlations of the low power data (transform limited pulse at 3KW) and the complete data that include all power levels is presented in Supplement 1 S3. It is worth noting that while the correlation function has the potential to display negative values (anti-correlations), none were present in these measurements. Experimentally measured values ranged from approximately 0 (indicating no correlations) to 1 (representing full correlations). The low-power case, Fig. 6(a), demonstrates a relatively homogeneous correlation pattern across all wavelengths. At the high powers, Fig. 6(b)-(d) a distinct correlation patterns emerge for each pulse that was used in the experiment. The transform-limited pulse shows high correlations at the spectral edges, while the down-chirp shows relatively high correlations around the central frequency. The up-chirped pulse shows relatively homogeneous correlation pattern across wavelengths.

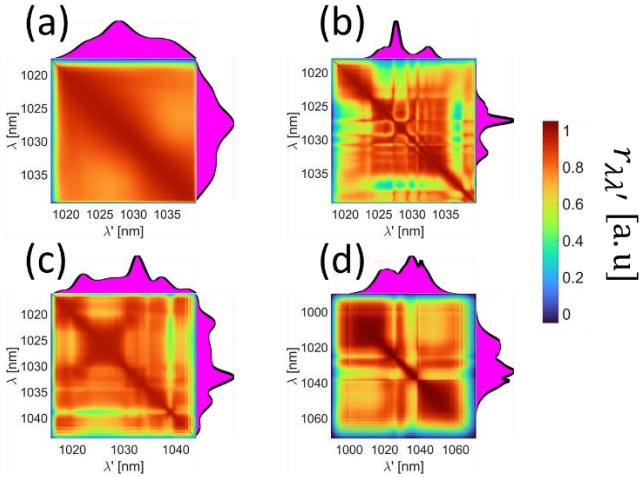

**Fig. 6.** Correlations for different chirped pulses with the spatially integrated spectrum depicted in pink. (a) Low power TL pulse (b) Down-chirped pulse (c) Up-chirped pulse (d) Transform-limited pulse.

In conclusion, we investigated the Kerr-induced self-cleaning phenomenon using transform limited and chirped femtosecond pulses. To gain deeper insights into the spatial-spectral characteristics of the cleaned beam, we employed a spectrally resolved imaging technique, which enabled us to detect and analyze complexities that would have otherwise remained unnoticed. We analyze the spatial-spectral homogeneity of the cleaned beam and demonstrate correlations between spectral broadening/narrowing of the pulse and homogeneity. Measurements show that for different chirped pulses, the beam cleanup distributes differently along the wavelengths, with beam homogeneity decreasing for all, with signs of saturations for down-chirped pulses. We also identified correlations between the spectral reshaping of the pulse and beam homogeneity. These findings significantly advance our understanding of Kerr-induced self-cleaning with chirped ultrafast pulses and offer new possibilities for diverse applications, such as generating arbitrarily complex fields, spatiotemporally mode-locked lasers, and high-power multimode amplifiers.

**Funding.** Eliyahu Pen Fund for Scientific and Medical Research.

**Disclosures.** The authors declare no conflict of interest.

**Data availability**. Data underlying the results presented in this paper are not publicly available at this time but may be obtained from the authors upon reasonable request.

**Supplemental document**. See Supplement 1 for supporting content.

**REFERENCES**

1. I. Cristiani, C. Lacava, G. Rademacher, *et al.*, J. Opt. **24**, 083001 (2022).
2. L. G. Wright, F. O. Wu, D. N. Christodoulides, *et al.*, Nat. Phys. **18**, 1018 (2022).
3. L. G. Wright, W. H. Renninger, D. N. Christodoulides, *et al.*, Optica **9**, 824 (2022).
4. W. H. Renninger and F. W. Wise, Nat. Commun. **4**, 1719 (2013).
5. K. Krupa, A. Tonello, A. Barthélémy, *et al.*, Phys. Rev. Lett. **116**, 183901 (2016).
6. L. G. Wright, D. N. Christodoulides, and F. W. Wise, Science **358**, 94 (2017).
7. L. G. Wright, D. N. Christodoulides, and F. W. Wise, Nat. Photonics **9**, 306 (2015).
8. K. Krupa, A. Tonello, B. M. Shalaby, *et al.*, Nat. Photonics **11**, 237 (2017).
9. Z. Liu, L. G. Wright, D. N. Christodoulides, *et al.*, Opt. Lett. **41**, 3675 (2016).
10. F. O. Wu, A. U. Hassan, and D. N. Christodoulides, Nat. Photonics **13**, 776 (2019).
11. F. Mangini, M. Gervaziev, M. Ferraro, *et al.*, Opt. Express **30**, 10850 (2022).
12. H. Pourbeyram, P. Sidorenko, F. O. Wu, *et al.*, Nat. Phys. **18**, 685 (2022).
13. Y. Leventoux, G. Granger, K. Krupa, *et al.*, Opt. Lett. **46**, 66 (2021).
14. K. Krupa, G. G. Castañeda, A. Tonello, *et al.*, Opt. Lett. **44**, 171 (2019).
15. Y. Leventoux, G. Granger, K. Krupa, *et al.*, Opt. Lett. **46**, 3717 (2021).
16. M. Hanna, F. Guichard, N. Daher, *et al.*, Laser & Photonics Reviews **15**, 2100220 (2021).
17. K. Pearson, Biometrika **13**, 25 (1920).

# Spatial-Spectral Complexity in Kerr Beam Self-Cleaning: Supplement

M. LABAZ[1, *] AND P. SIDORENKO[2].

[1] Department of Physics and Solid-State Institute, Technion – Israel Institute of Technology, 32000 Haifa, Israel
[2] Department of Electrical and Computer Engineering and Solid-State Institute, Technion – Israel Institute of Technology, 32000 Haifa, Israel
*Corresponding author: l.moshe@campus.technion.ac.il

## S1. System calibration.

In the experimental protocol, we calibrated the scanning setup by the following procedure: a) we generate a highly-multimode speckle pattern by uniformly illuminating the fiber input. In this case, for a multimode fiber the near-field pattern approximates the refractive index profile of the fiber core. b) We scan the image plane with the single mode fiber and integrate over the wavelengths. We use this approach to find the fiber center and the core diameter. We perform this for multiple speckle patterns (an example of one such image is shown in Fig. S1) and use the average value for the fiber center and the core diameter.

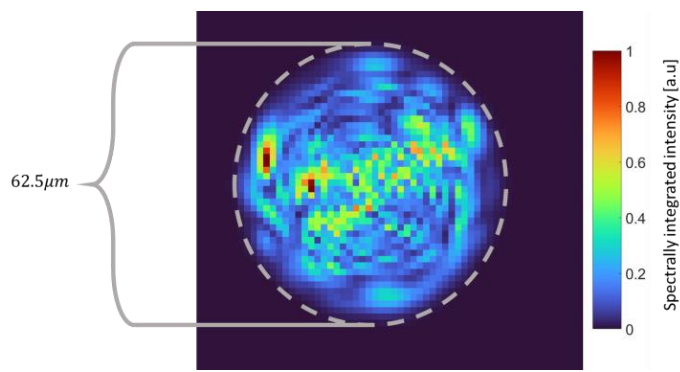

**Fig. S1** Spectrally integrated spatial picture of many modes excited GRIN MMF with emphasizing the actual core size of - $62.5 \mu m$ in diameter.

## S2. Spatial-spectral results for all the measurements:

Below are the full spatial-spectral data that was measured in our experiments. Figures S2, S3 and S4 show the measured data for the transform limited, up-chirped and down-chirped pulses respectively. Each figure shows the spectrally integrated output beam profile (top image on each figure with marked corresponding peak power). The middle row in each figure shows the spatially integrated power spectrum and exemplary beam profiles at several wavelengths. The bottom row in each figure shows an isosurfaces of the measure data $I(x,y,\lambda)$ and corresponding projection on x-$\lambda$, y-$\lambda$ and x-y planes.

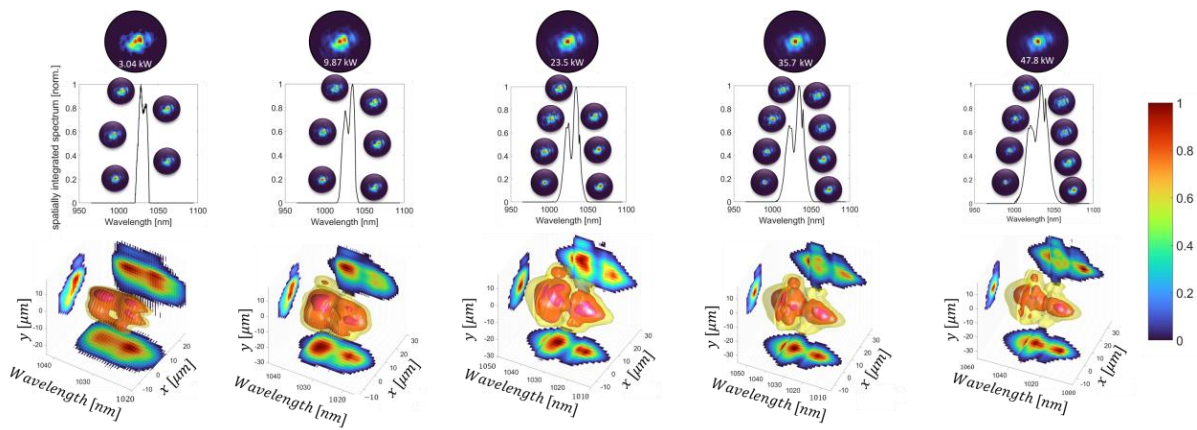

**Fig S2** Full spatial-spectral images for the Transform-limited pulse for different peak powers.

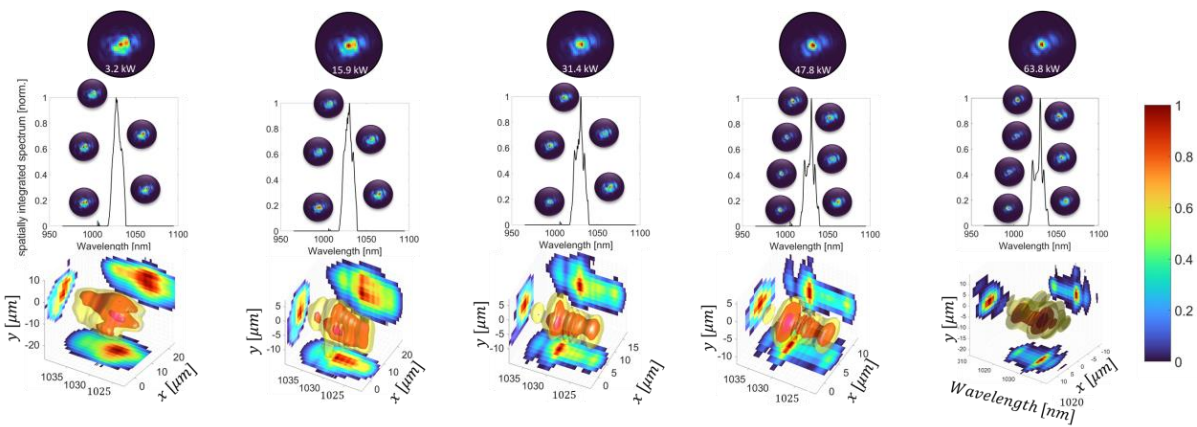

**Fig S3** Full spatial-spectral images for the Up-chirped pulse for different peak powers.

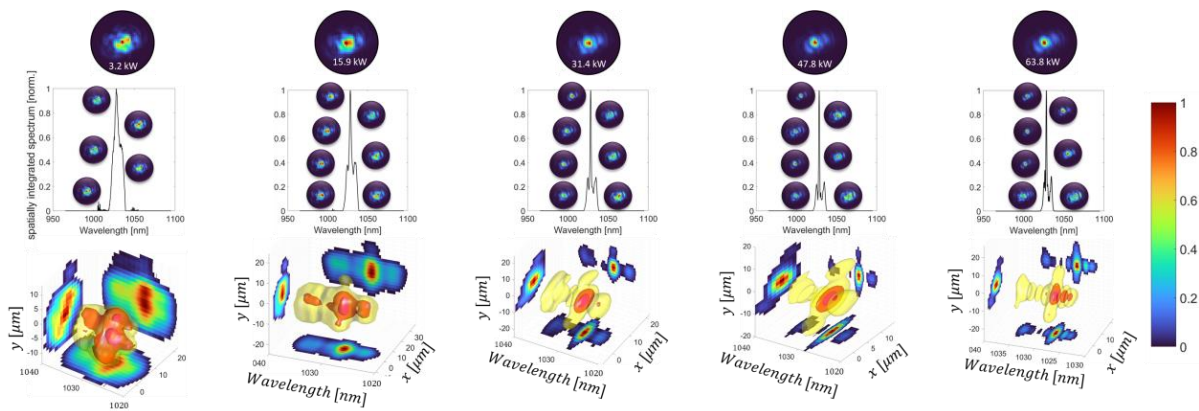

**Fig S4** Full spatial-spectral images for the Down-chirped pulse for different peak powers.

## S3. Pulse correlations for all the measurements:

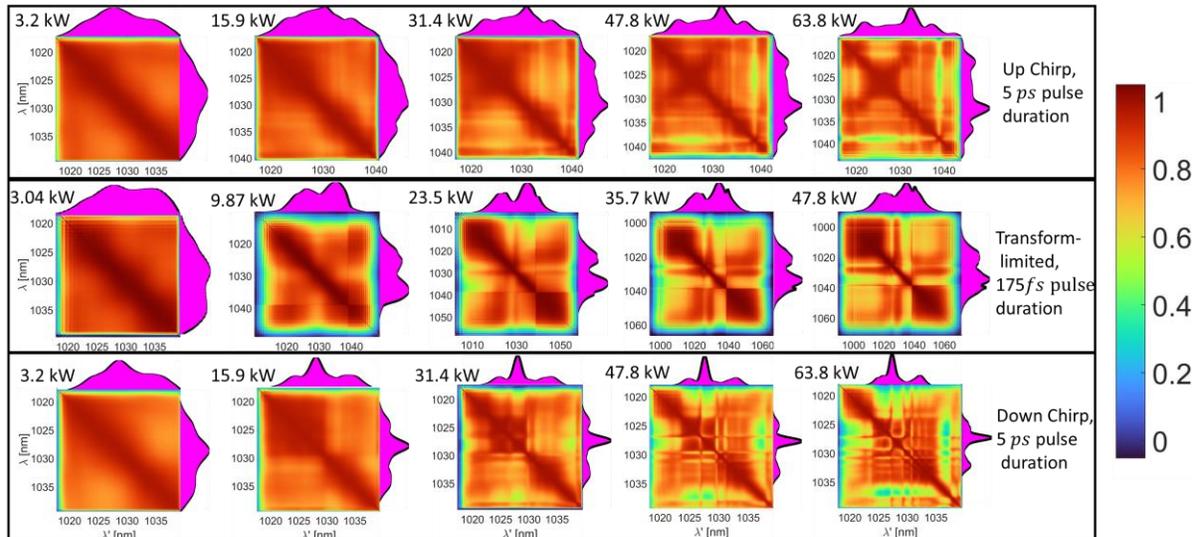

**Fig S. 3** Correlations for all the chirped pulses (first and third rows) and for transform-limited pulse (second row) with spatially integrated spectral shapes (depicted in pink)

## Full Reference list

1. I. Cristiani, C. Lacava, G. Rademacher, B. J. Puttnam, R. S. Luìs, C. Antonelli, A. Mecozzi, M. Shtaif, D. Cozzolino, D. Bacco, L. K. Oxenløwe, J. Wang, Y. Jung, D. J. Richardson, S. Ramachandran, M. Guasoni, K. Krupa, D. Kharenko, A. Tonello, S. Wabnitz, D. B. Phillips, D. Faccio, T. G. Euser, S. Xie, P. S. J. Russell, D. Dai, Y. Yu, P. Petropoulos, F. Gardes, and F. Parmigiani, "Roadmap on multimode photonics," J. Opt. **24**(8), 083001 (2022).
2. L. G. Wright, F. O. Wu, D. N. Christodoulides, and F. W. Wise, "Physics of highly multimode nonlinear optical systems," Nature Physics **18**(9), 1018–1030 (2022).
3. L. G. Wright, W. H. Renninger, D. N. Christodoulides, and F. W. Wise, "Nonlinear multimode photonics: nonlinear optics with many degrees of freedom," Optica **9**(7), 824–841 (2022).
4. W. H. Renninger and F. W. Wise, "Optical solitons in graded-index multimode fibres," Nat Commun **4**(1), 1719 (2013).
5. K. Krupa, A. Tonello, A. Barthélémy, V. Couderc, B. M. Shalaby, A. Bendahmane, G. Millot, and S. Wabnitz, "Observation of Geometric Parametric Instability Induced by the Periodic Spatial Self-Imaging of Multimode Waves," Phys. Rev. Lett. **116**(18), 183901 (2016).
6. L. G. Wright, D. N. Christodoulides, and F. W. Wise, "Spatiotemporal mode-locking in multimode fiber lasers," Science **358**(6359), 94–97 (2017).
7. L. G. Wright, D. N. Christodoulides, and F. W. Wise, "Controllable spatiotemporal nonlinear effects in multimode fibres," Nature Photon **9**(5), 306–310 (2015).
8. K. Krupa, A. Tonello, B. M. Shalaby, M. Fabert, A. Barthélémy, G. Millot, S. Wabnitz, and V. Couderc, "Spatial beam self-cleaning in multimode fibres," Nature Photon **11**(4), 237–241 (2017).
9. Z. Liu, L. G. Wright, D. N. Christodoulides, and F. W. Wise, "Kerr self-cleaning of femtosecond-pulsed beams in graded-index multimode fiber," Opt. Lett., OL **41**(16), 3675–3678 (2016).
10. F. O. Wu, A. U. Hassan, and D. N. Christodoulides, "Thermodynamic theory of highly multimoded nonlinear optical systems," Nature Photonics **13**(11), 776–782 (2019).
11. F. Mangini, M. Gervaziev, M. Ferraro, D. S. Kharenko, M. Zitelli, Y. Sun, V. Couderc, E. V. Podivilov, S. A. Babin, and S. Wabnitz, "Statistical mechanics of beam self-cleaning in GRIN multimode optical fibers," Optics Express **30**(7), 10850 (2022).
12. H. Pourbeyram, P. Sidorenko, F. O. Wu, N. Bender, L. Wright, D. N. Christodoulides, and F. Wise, "Direct observations of thermalization to a Rayleigh–Jeans distribution in multimode optical fibres," Nat. Phys. **18**(6), 685–690 (2022).
13. Y. Leventoux, G. Granger, K. Krupa, A. Tonello, G. Millot, M. Ferraro, F. Mangini, M. Zitelli, S. Wabnitz, S. Février, and V. Couderc, "3D time-domain beam mapping for studying nonlinear dynamics in multimode optical fibers," Opt. Lett., OL **46**(1), 66–69 (2021).
14. K. Krupa, G. G. Castañeda, A. Tonello, A. Niang, D. S. Kharenko, M. Fabert, V. Couderc, G. Millot, U. Minoni, D. Modotto, and S. Wabnitz, "Nonlinear polarization dynamics of Kerr beam self-cleaning in a graded-index multimode optical fiber," Opt. Lett., OL **44**(1), 171–174 (2019).
15. Y. Leventoux, G. Granger, K. Krupa, T. Mansuryan, M. Fabert, A. Tonello, S. Wabnitz, V. Couderc, and S. Février, "Frequency-resolved spatial beam mapping in multimode fibers: application to mid-infrared supercontinuum generation," Opt. Lett., OL **46**(15), 3717–3720 (2021).
16. M. Hanna, F. Guichard, N. Daher, Q. Bournet, X. Délen, and P. Georges, "Nonlinear Optics in Multipass Cells," Laser & Photonics Reviews **15**(12), 2100220 (2021).
17. K. Pearson, "Notes on the history of correlation," Biometrika **13**(1), 25–45 (1920).